\documentclass[11pt,aps,prd,preprint,superscriptaddress]{revtex4-1}
\usepackage{amsmath}
\usepackage[dvips]{graphicx}
\usepackage[english]{babel}
\usepackage{wasysym}

\begin{document}
\title{Three thermodynamically-based parametrizations of the deceleration parameter}
\author{Sergio del Campo\footnote{E-mail: sdelcamp@ucv.cl}}
\affiliation{Instituto de F\'{\i}sica, Pontificia Universidad
Cat\'{o}lica de Valpara\'{\i}so, Av. Universidad 330, Campus
Curauma, Valpara\'{\i}so, Chile}
\author{Ivan Duran\footnote{E-mail: ivan.duran@uab.cat}}
\affiliation{Departamento de F\'{\i}sica, Facultad de Ciencias,
Universidad Aut\'{o}noma de Barcelona, 08193 Bellaterra
(Barcelona), Spain}
\author{Ram\'{o}n Herrera\footnote{E-mail: ramon.herrera@ucv.cl}}
\affiliation{Instituto de F\'{\i}sica, Pontificia Universidad
Cat\'{o}lica de Valpara\'{\i}so, Av. Universidad 330, Campus
Curauma, Valpara\'{\i}so, Chile}
\author{Diego Pav\'{o}n\footnote{E-mail: diego.pavon@uab.es}}
\affiliation{Departamento de F\'{\i}sica, Facultad de Ciencias,
Universidad Aut\'{o}noma de Barcelona, 08193 Bellaterra
(Barcelona), Spain}
\begin{abstract}
We propose,  and constrain with the latest observational data,
three model-independent parametrizations of the cosmic
deceleration parameter $q(z)$. They are well behaved and stay
finite at all redshifts. We construct them by fixing the value of
$q$ at high redshift, $q(z \gg 1) = 1/2$ (as demanded by cosmic
structure formation), and at the far future, $q(z = -1) = -1$, and
smoothly interpolating $q(z)$ between them. The fixed point at $z
= -1$ is not arbitrarily chosen; it readily follows from the
second law of thermodynamics. This fairly reduces the ample
latitude in parameterizing $q(z)$.
\end{abstract}

\maketitle
\section{Introduction}\label{sec:introduction}
The deceleration parameter, defined as $q = -1 -(\dot{H}/H^{2})$,
is a key quantity in describing the  evolution  of the homogeneous
and isotropic universe. Its importance lies in the fact that it
tells us the rate at which the Universe accelerates or decelerates
its expansion. (Here and throughout $H = \dot{a}/a$ denotes the
Hubble function, and $a$ the scale factor of the
Friedmann-Lema\^{i}tre-Robertson-Walker (FLRW) metric).
Unfortunately, at present measurements of $q$ suffer from
non-small uncertainties that quickly grow with redshift ($z =
a^{-1}-1$ with $a_{0} = 1$), though it is virtually certain that
the Universe is accelerating nowadays, i.e., that $q_{0} <0 $ (the
zero subscript means present time). Expressions of $q(z)$ provided
by cosmological models are of not much help either because none of
them  rests on sufficiently convincing theoretical grounds. It is
to be hoped that things will eventually turn for the better when a
successful theory of quantum gravity is in place, though it may
well take a long while.

Nevertheless, on the observational side the situation may improve
comparatively soon given the variety and range of ongoing and
planned major ambitious projects that involve bigger telescopes
and advanced techniques -for a short review and a comprehensive
list of references see \S $14$ of \cite{projects}. In view of the
above, it seems reasonable to propose parametrized expressions of
$q(z)$ based not in any specific cosmological model but on
practical and empirical reasons that lessen their wide latitude.
They can be useful while we patiently wait for a theoretically
sound model backed by observation at all scales.

Thus far, different parametrizations, such as $q = q_{0} \, + \,
q_{1} z$, $q = q_{0} \, + \, q_{1} z (1 \, + \, z)^{-1}$, $q =
q_{1} \, +\, q_{2}z(1+z)^{-2}$,  $q = 1/2 \, + \,
q_{1}(1+z)^{-2}$, $q = 1/2 \, + \, (q_{1}z \, + \,
q_{2})(1+z)^{-2}$, and more complex than these, have been
considered in the literature to reconstruct $q(z)$ from
observational data (see e.g. \cite{elgaroy}-\cite{nair}). However,
the first parametrization is adequate for $\mid z \mid \ll 1$ only
and the others are unsuitable to predict the behavior of the
deceleration parameter in the far future; in particular, they
diverge as $z \rightarrow -1$. Parametrizations whose intended
range of validity includes the far future are necessarily more
involved and usually contain three or four free parameters
\cite{ishida,giostri}.

In this paper  we propose three model independent
parametrizations, with just two free parameters, valid from the
matter dominated epoch ($z \gg 1$) onwards (i.e., up to $z = -1$),
based on practical and theoretical reasons and independent of any
cosmological model. By construction they obey the asymptotic
conditions, $q(z \gg 1) = 1/2$, $q(z = -1) = -1$, and a further
condition, $dq/dz > 0$, which is valid at least when $q
\rightarrow -1$. The first condition expresses the conviction that
at sufficiently high redshift the Universe was matter dominated
(otherwise it would be very hard to account for the observed
cosmic structures). At first sight, the other conditions are less
compelling. As explained below, they are based on the second law
of thermodynamics when account is made of the entropy of the
apparent horizon. Usually one parametrizes a function in any
specific interval by interpolating it between two given points
(one at either end of the interval), modulo one first knows the
value taken by the function at these two points. In actual fact,
the parametrizations of $q(z)$ proposed so far have just one fixed
point: the asymptotic value at high redshift ($q$ must converge to
$1/2$ when $z \gg 1$). The other, $q_{0}$, is not in reality a
fixed point  because the value of the deceleration parameter at $z
= 0$ is not very well known and therefore left free. The
parametrizations proposed in this paper have two fixed points, one
at the far past ($z \gg1$), and other at the far future ($z=-1$).
The second fixed point conforms to the thermodynamical constraints
imposed by the second law. We believe this means a clear advantage
over previous parametrizations of $q(z)$, with just one fixed
point. While it can be found in the literature parameterizations
that also fix $q$ at $z = -1$ they do so arbitrarily, i.e., not
grounded on sound physics.

The aim of this paper is  to propose three model-independent
parametrizations of $q(z)$, from $z \gg 1$ up to $z = -1$, that
comply with the second law of thermodynamics and constrain their
two free parameters with recent observational data. As it turns
out, all of them predict that the present stage of accelerated
expansion will never slow down and are consistent with the
spatially flat $\Lambda$CDM model.

Section \ref{sec:thermo} considers the properties the deceleration
parameter must observe assuming the Universe obeys the second law
of thermodynamics. Put otherwise if, in the long run, it tends to
an equilibrium state, i.e., of maximum entropy. Section
\ref{sec:statistics} presents the statistical tools, to be
employed later, that make use of observational data from
supernovae type Ia (557 data points), baryon acoustic oscillations
combined with cosmic microwave background (BAO/CMB) (7 data
points), and the history of the Hubble factor (24 data points).
Section \ref{sec:parm} presents the three parametrizations,
constrain them with the observational data, and discuss them.
Finally, Section \ref{sec:remarks} summarizes our work and
introduces some final remarks.

\section{Thermodynamical constraints on $q(z)$}\label{sec:thermo}
As is well known, physical systems tend spontaneously to some
equilibrium state compatible with the constraints imposed on them.
This summarizes the empirical basis of the second law of
thermodynamics. Very briefly, this law establishes that isolated,
macroscopic systems, evolve to the maximum entropy state
consistent with their constraints \cite{callen}. As a consequence
their entropy, $S$, cannot decrease at any time, i.e., $dS \geq
0$. Further, in the last phase of the evolution $S$ has to be a
convex function of the  said variable, $d^{2}S(z\rightarrow -1) <
0$.

Arguably,  the entropy of Friedmann-Lema\^{i}tre-Robertson-Walker
(FLRW) universes is dominated by the entropy of the causal
horizon, at least at late times -see e.g. \cite{egan}. As causal
horizon we shall take the apparent horizon, the marginally trapped
surface with vanishing expansion of radius $\tilde{r}_{A} =
1/\sqrt{H^{2}\, + \, k \, a^{-2}}\, $ \cite{bak-rey}, where $k$
denotes  the spatial curvature index. Interestingly enough, it has
been shown that this horizon represents the appropriate
thermodynamic boundary surface \cite{boundary}. Leaving aside
possible quantum corrections its entropy results proportional to
area of the latter ($4\pi \tilde{r}_{A}^{2}$) \cite{bak-rey},
\begin{equation}
S_{A} \propto {\cal A} = 4\pi \, (H^{2}\, + \, k\, a^{-2})^{-1}\,
. \label{SA}
\end{equation}
Therefore, so long as we can ignore the entropy within the
horizon, the second law of thermodynamics imposes ${\cal A}' \geq
0$, at any time, as well as ${\cal A}'' \leq 0$ at late times -the
prime meaning derivative with respect to the scale factor. Both
conditions are to be fulfilled if the FLRW universe is to tend to
thermodynamic equilibrium at late times \cite{nd2010}.

Bearing in mind  the definition of the deceleration parameter, for
spatially flat ($k = 0$) FLRW universes we can write
\begin{equation}
{\cal A}' = 2 {\cal A} \, \frac{1+q}{a} \, , \qquad {\rm and}
\qquad {\cal A}'' = 2 {\cal A}\, \left[ 2
\left(\frac{1+q}{a}\right)^{2}\, + \, \frac{q'}{a} \, - \, 2 \,
\frac{1+q}{a^{2}}\right] \, . \label{Aprimes}
\end{equation}
The first equation implies $q \geq -1$. Inspection of the second
one reveals that when $a \rightarrow \infty$ the middle term in
the square parenthesis dominates. Thereby, $dq/da <0$ in that
limit. Thus, for the Universe to tend to thermodynamic equilibrium
at late times we must have $q \rightarrow -1$ and $dq/dz > 0$ as
$z \rightarrow -1$.

\section{Statistical tools}\label{sec:statistics}
This Section outlines the use of the observational data to fit the
parametrizations of the deceleration parameter, $q(z)$. Since the
likelihood function is defined by $ {\cal L} \propto \exp (-
\chi^{2}/2)$ the best fit to the data follows from minimizing the
sum $\chi^{2}_{\rm total} = \chi^{2}_{SN} \, + \,
\chi^{2}_{BAO/CMB} \, +  \, \chi^{2}_{H}$. As detailed below, the
best fit values of the parameters can be obtained by contrasting
the proposed parametrizations with the empirical data mentioned
above and  minimizing  the $\chi^{2}_{\rm total}$ by means of the
Markov Chain Monte Carlo method.

\subsection {SN Ia}
We compare the theoretical distance modulus
\begin{equation}
 \mu_{th}(z_{i}) = 5\log_{10}\left(\frac{d_{L}}{{10{\rm
pc}}}\right)\, + \,\mu_{0}\, , \label{modulus}
\end{equation}
where $\mu_{0} = 42.38 \, -\, 5\log_{10} h$, with the observed
distance modulus $\mu_{obs}(z_{i})$ of the 557 supernovae type Ia
assembled in the Union2 compilation \cite{Amanullah}. The latter
data set is substantially richer than previous SN Ia compilations
and presents other advantages; mainly, the refitting of all light
curves with the SALT2 fitter and an upgraded control of systematic
errors. In (\ref{modulus}) $d_{L} = (1+z)
\int_{0}^{z}{\frac{d\tilde{z}}{E(\tilde{z};{\bf p})}}$ is the
Hubble-free luminosity distance, ${\bf p} \equiv (q_{1},q_{2})$,
and  $E(z;{\bf p}) = H(z;{\bf p})/H_{0}$. Here and throughout,
$q_{1}$ and $q_{2}$ denote the free parameters occurring in the
parametrizations presented below in Section IV; $H_{0}$ is the
Hubble constant and $h$ its value in units of $100$ km/s/Mpc.

The $\chi^{2}$ from the 557 SN Ia is given by
\begin{equation}
 \chi^{2}_{SN}({\bf p}) = \sum_{i=1}^{557}
\frac{[\mu_{th}(z_{i}) \, - \,
\mu_{obs}(z_{i})]^{2}}{\sigma^{2}(z_{i})} \, , \label{chi2mu}
\end{equation}
where the subscripts ``$th$" and ``$obs$" indicate the theoretical
value (i.e., the value from the parametrization) and the observed
value, respectively. As usual, the $\sigma_{i}$ quantities stand
for the 1$\sigma$ uncertainty associated to the $i$th data point.
To eliminate the effect of the nuisance parameter $\mu_{0}$, which
is independent of the data points and the data set, we follow the
procedure of \cite{Nesseris-Perilovorapoulos}.
\subsection{BAO and CMB}
Baryon acoustic oscillations can be traced to pressure waves at
the recombination epoch generated by cosmological perturbations in
the primeval baryon-photon plasma. They have been revealed by a
distinct peak in the large scale correlation function measured
from the luminous red galaxies sample of the Sloan Digital Sky
Survey (SDSS):  at $z = 0.35$ \cite{Eisenstein}, as well as in the
two degree Field Galaxy Redshift Survey  at $z = 0.2$
\cite{Percival}. More recently other peaks have been observed: at
$z=0.278$ (with the SDSS \cite{Kazin}), at $z=0.106$ (in the six
degree Field Galaxy Redshift Survey \cite{Beutler}), and at
$z=0.44$, $z=0.60$, and $z=0.73$ (by the WiggleZ team
\cite{Blake}).

\noindent From each peak the ratio of the comoving sound horizon $
r_{s}(z) =\int_{z}^{\infty}{c_{s}(z)/H(z)\, dz}$ at decoupling ($z
= z_{\star} \simeq 1090$) and at the drag epoch ($z = z_{d}$), the
epoch at which the acoustic oscillations are frozen in, can be
measured.
Here $c_{s}$ is the speed of sound. Likewise, at each peak, a
characteristic distance scale, the dilation scale
\begin{equation}\label{eq:BAO}
D_{V}(z_{BAO})=\left[z_{BAO}\frac{d_{A}^{2}(z_{BAO})}{H(z_{BAO})}
\right]^{\frac{1}{3}} \, ,
\end{equation}
\noindent where $d_{A}(z)=\int_{0}^{z}\frac{dz'}{H(z')}$ is the
comoving angular distance, can also be determined.
To compute the drag epoch redshift we use the formula $(4)$ of
Eisenstein and Hu in \cite{EisensteinHu} and get $z_{d}\approx
1020$.

Multiplying the ratio, $\frac{r_{s}(z_{d})}{D_{V}(z_{BAO})}$,
taken from the BAO peaks by the acoustic scale
\begin{equation}\label{eq:lA}
l_{A}=\pi\frac{d_{A}(z_{\star})}{r_{s}(z_{\star})}\, ,
\end{equation}
we get
$\frac{d_{A}(z_{\star})}{D_{V}(z_{BAO})}\frac{r_{s}(z_{d})}{r_{s}(z_{\star})}$
at each redshift of the seven BAO data. Here we use the value for
$l_{A}$ derived from Wilkinson microwave anisotropy probe
(WMAP)7-years data, namely, $l_{A}=302.09 \pm 0.76$
\cite{komatsu}. If we also use the value of the ratio of sound
horizon at the drag epoch and at recombination (redshift
$z_{\star}$), computed from the values reported in \cite{komatsu}
$\frac{r_{s}(z_{d})}{r_{s}(z_{\star})}=1.045 \pm 0.015$ we obtain
the new estimator $\frac{d_{A}(z_{\star})}{D_{V}(z_{BAO})}$, shown
in table \ref{tab:BAOCMB}, as done in \cite{Sollerman}. Using this
estimator, the dependence in the sound horizons at decoupling and
the drag epoch is suppressed. Thus we just use the ratio between
them, which is almost model independent. This follows because both
redshifts are rather close and the sound horizon at decoupling and
drag essentially depend on the fractional difference between the
number of photons and baryons \cite{Sollerman}.
\begin{table}[!htb]\scriptsize
    \begin{tabular}{||p{1.5 cm} || p{2 cm} p{2 cm} p{2 cm} p{2 cm} p{2 cm} p{2 cm} p{2 cm}||}
    \hline \hline
    $z_{BAO}$ &$ \qquad \, 0.106$ &  $ \qquad \; 0.2$ &  $\qquad \; 0.278$ & $\qquad \; 0.35$ & $\qquad \; 0.44$ & $\qquad \; 0.6$\qquad \;  & $\qquad \; 0.73$\\
    \hline \hline
    $\frac{r_{s}(z_{d})}{D_{V}(z_{BAO})}$&$\; 0.336 \pm 0.015$  & $0.1905 \pm 0.0061$ & $0.1394 \pm 0.0049$ & $0.1097 \pm 0.0036$ & $0.0916 \pm 0.0071$ & $0.0726 \pm 0.034$ & $0.0592 \pm 0.0032$\\
    \hline
    $\frac{d_{A}(z_{\star})}{D_{V}(z_{BAO})}$ &$\; 30.92 \pm 1.45$  & $17.53 \pm 0.62$ & $12.83 \pm 0.49$ & $10.09 \pm 0.36$ & $8.43 \pm 0.66$ & $6.68 \pm 0.33$ & $5.45 \pm 0.30$\\
    \hline \hline
    \end{tabular}
    \caption{Values of $\frac{r_{s}(z_{d})}{D_{V}(z_{BAO})}$
    (reported in \cite{Eisenstein,Percival,Beutler,Kazin,Blake}) and the
    derived ratio $\frac{d_{A}(z_{\star})}{D_{V}(z_{BAO})}$.}
    \label{tab:BAOCMB}
\end{table}
To obtain the $\chi^{2}$ for the combined BAO/CMB data we compute
\begin{equation}\label{eq:ChiBAOCMB}
\chi^{2}_{BAO/CMB}=\textbf{X}^{T}\textbf{C}^{-1}\textbf{X}\, ,
\end{equation}
\noindent where
\begin{equation*}\label{eq:XBAOCMB}
\textbf{X}=
\begin{pmatrix}
\frac{d_{A}(z_{\star})}{D_{V}(0.106)} - 30.92\\
\frac{d_{A}(z_{\star})}{D_{V}(0.2)} - 17.53\\
\frac{d_{A}(z_{\star})}{D_{V}(0.278)} - 12.83\\
\frac{d_{A}(z_{\star})}{D_{V}(0.35)} - 10.09\\
\frac{d_{A}(z_{\star})}{D_{V}(0.44)} - 8.43\\
\frac{d_{A}(z_{\star})}{D_{V}(0.6)} - 6.68\\
\frac{d_{A}(z_{\star})}{D_{V}(0.73)} - 5.45\\
\end{pmatrix}
\end{equation*}
\noindent and $\textbf{X}^{T}$ the transpose matrix. The elements
of covariance matrix $\textbf{C}$ are given by
\begin{equation}\label{eq:CovarianceMatrix}
\textbf{C}_{ij}=\sum_{k}\left(\frac{\partial
\frac{d_{A}(z_{\star})}{D_{V}(z)}}{\partial p_{k}}\right)_{z_{i}}
\left(\frac{\partial \frac{d_{A}(z_{\star})}{D_{V}(z)}}{\partial
p_{k}}\right)_{z_{j}}\textbf{C}_{p_{k}\;i\;j} \, ,
\end{equation}
\noindent where the sum is over the estimators used (in our case,
$\frac{r_{s}(z_{d})}{D_{V}(z_{BAO})}$, $l_{A}/\pi$ and
$\frac{r_{s}(z_{d})}{r_{s}(z_{\star})}$). The elements of the
original covariance matrices are
$\textbf{C}_{\frac{l_{A}}{\pi}}=\theta_{\frac{l_{A}}{\pi}}^{2}$,
$\textbf{C}_{\frac{r_{s}(z_{d})}{r_{s}(z_{\star})}}=
\theta_{\frac{r_{s}(z_{d})}{r_{s}(z_{\star})}}^{2}$ and
$\textbf{C}_{BAO\;i\,j}=\theta_{BAO\;i}\; \, \theta_{BAO\;j}\; \,
r_{i\,j}$ where $\theta_{BAO\;i}$ stand for the errors associated
with the estimator $\frac{r_{s}(z_{d})}{D_{V}(z_{i})}$. The only
non-zero off-diagonal correlation coefficients $r_{i\,j}$ are
$r_{z=0.2\,z=0.35}=0.337$, $r_{z=0.44\,z=0.6}=0.369$ and
$r_{z=0.6\,z=0.73}=0.438$, and their symmetric
\cite{Percival,Blake}. Thus, the inverse covariance matrix comes
to be
\begin{equation*}\label{eq:InvCovarianceBAOCMB}
\textbf{C}^{-1}=
\begin{pmatrix}
0.492 & -0.084 & -0.126 & -0.136 & -0.025 & -0.081 & -0.088\\
-0.084 & 3.362 & -0.327 & -2.397 & -0.065 & -0.209 & -0.228\\
-0.126 & -0.327 & 4.429 & -0.528 & -0.098 & -0.314 & -0.342\\
-0.136 & -2.397 & -0.528 & 9.712 & -0.106 & -0.338 & -0.368\\
-0.025 & -0.065 & -0.098 & -0.106 & 2.798 & -2.749 & 1.182\\
-0.081 & -0.209 & -0.314 & -0.338 & -2.749 & 15.002 & -7.294\\
-0.088 & -0.228 & -0.342 & -0.368 & 1.182 & -7.294 & 14.587\\
\end{pmatrix}\, .
\end{equation*}
\subsection{History of the Hubble parameter}
The history of the Hubble parameter, $H(z)$, is poorly constrained
though, recently, some high precision measurements by Riess {\em
et al.} at $z = 0$, obtained from the observation of 240 Cepheid
variables of rather similar periods and metallicities
\cite{Riess-2009}, and Gazta\~{n}aga {\em et al.} at $z = 0.24, \,
0.34, {\rm and}\,  0.43 \,$ \cite{gazta}, who used the BAO peak
position as a standard ruler in the radial direction, have
improved matters somewhat. We have employed these four data
alongside 11 less precise data, in the redshift interval $0.1
\lesssim z \lesssim 1.8$, from Simon {\em et al.} \cite{simon} and
Stern {\em et al.} \cite{stern}, derived from the differential
ages of passive-evolving galaxies and archival data. In addition
we have  included in our analysis 9 more recent correlated data
from the WiggleZ survey \cite{Blake_Hz}.

The corresponding $\chi^{2}$
\begin{equation}
 \chi^{2}_{H}({\bf p}) = \sum_{i=1}^{15}
\frac{[H_{th}(z_{i}) \, - \,
H_{obs}(z_{i})]^{2}}{\sigma^{2}(z_{i})} \,
+\textbf{X}_{H}^{T}\textbf{C}_{H}^{-1}\textbf{X}_{H}\, \, ,
 \label{chi2hubble}
 \end{equation}
\noindent where
\begin{equation*}\label{eq:XH}
\textbf{X}_{H}=
\begin{pmatrix}
    H_{th}(0.05) - 69.4\\
    H_{th}(0.15) - 76.6\\
    H_{th}(0.25) - 75.3\\
    H_{th}(0.35) - 78.3\\
    H_{th}(0.45) - 87.3\\
    H_{th}(0.55) - 88.9\\
    H_{th}(0.65) - 101.4\\
    H_{th}(0.75) - 96.9\\
    H_{th}(0.85) - 127.3\\
\end{pmatrix}
\end{equation*}
\noindent and $\textbf{C}_{H}^{-1}$ is the inverse covariance
matrix given in table 6 of \cite{Blake_Hz}.
\section{Parametrizations}\label{sec:parm}
Here we propose and constrain three parametrizations of the
deceleration parameter, valid from the matter dominated era up to
$z = -1$. These fulfill: $(i)$ $q(z \gg 1) = 1/2$ (as demanded by
cosmic structure formation), $(ii)$ $q(z = -1) = -1$ and $dq(z)/dz
>0$ when $q(z) \rightarrow -1$ as required by the thermodynamic
arguments of above (the second law of thermodynamics). In
interpolating between  $(z = -1, \; q = -1)$ and $(z \gg 1, \; q=
1/2)$  we introduce two free parameters $q_{1}$ and $q_{2}$ and
fit them to the observational sets of data by the method of last
Section. Note that due to the scarcity of $q(z)$ data and their
big error bars, we do not constrain the parametrizations directly.
We constrain instead the expressions for $H(z)$ that arise from
integrating them; namely,
\begin{equation}
H(z) = H_{0} \, \exp\left\{\int_{0}^{z}{[1+q(x)] \, d \ln (1+x)}
\right\} \, , \label{eq:H(z)general}
\end{equation}
which holds for all parametrizations. This has the advantage of a
much bigger and robust statistics. Notice that the Hubble constant
also enters this expression as a free parameter. Its value for
each parametrization is obtained by fitting it to the $H(z)$ data,
\cite{Riess-2009,gazta,simon,stern,Blake_Hz}.


\subsection{Parametrization I}
As a first parametrization we propose
\begin{equation}
\label{eq:q1} q(z)
=-1+\frac{3}{2}\left(\frac{(1+z)^{q_{2}}}{q_{1}+(1+z)^{q_{2}}}\right)\,
,
\end{equation}
where to avoid divergences $q_{1}$ and $q_{2}$ must be
positive-definite.

Introducing (\ref{eq:q1}) in (\ref{eq:H(z)general}), it follows
\begin{equation}
H(z) =
 H_{0}\left(\frac{q_{1}+(1+z)^{q_{2}}}{q_{1}+1}\right)^{\frac{3}{2q_{2}}}
\, .
\label{eq:H1}
\end{equation}
(We note in passing that for  $q_{2}=3$ the $\Lambda$CDM behavior
is reproduced). By using the method outlined in last section in
conjunction with the observational data (SN Ia (557), CMB/BAO (7)
and $H(z)$ (15)), we fit the three free parameters occurring in
(\ref{eq:H1}). The result is $q_{1}=2.87^{+0.70}_{-0.53}$,
$q_{2}=3.27\pm 0.55$, and $H_{0}=70.5^{+1.5}_{-1.6}$ km/s/Mpc.
Table \ref{tab:param1} shows the $\chi^{2}$ values of the best
fit.
\begin{table}[!htb]
    \begin{tabular}{||p{5 cm} || p{1.5 cm} p{2.0 cm} p{1.6 cm} p{1.8 cm} p{1.8 cm}||}
    \hline \hline
    Data sets &$\; \chi^{2}_{SN}$ &  $\chi^{2}_{BAO/CMB}$&  $ \quad \; \chi^{2}_{H}$ &  $ \; \chi^{2}_{{\rm tot}}$ & $\chi^{2}_{{\rm tot}}/dof$\\
    \hline \hline
    \footnotesize{Union2+BAO/CMB+Hubble} &$\; 542.6$  & $\quad \; \; 2.6$ & $\quad \;17.9$ &  $563.3$ & $\; \; \; 0.96$ \\
    \hline \hline
    \end{tabular}
    \caption{Best fit $\chi^{2}$ values of parametrization I, Eq. (\ref{eq:q1}). The free parameters
    are $q_{1}$, $q_{2}$ and $H_{0}$.}
    \label{tab:param1}
\end{table}

\begin{figure}[!htb]
  \begin{center}
    \begin{tabular}{cc}
      \resizebox{80mm}{!}{\includegraphics{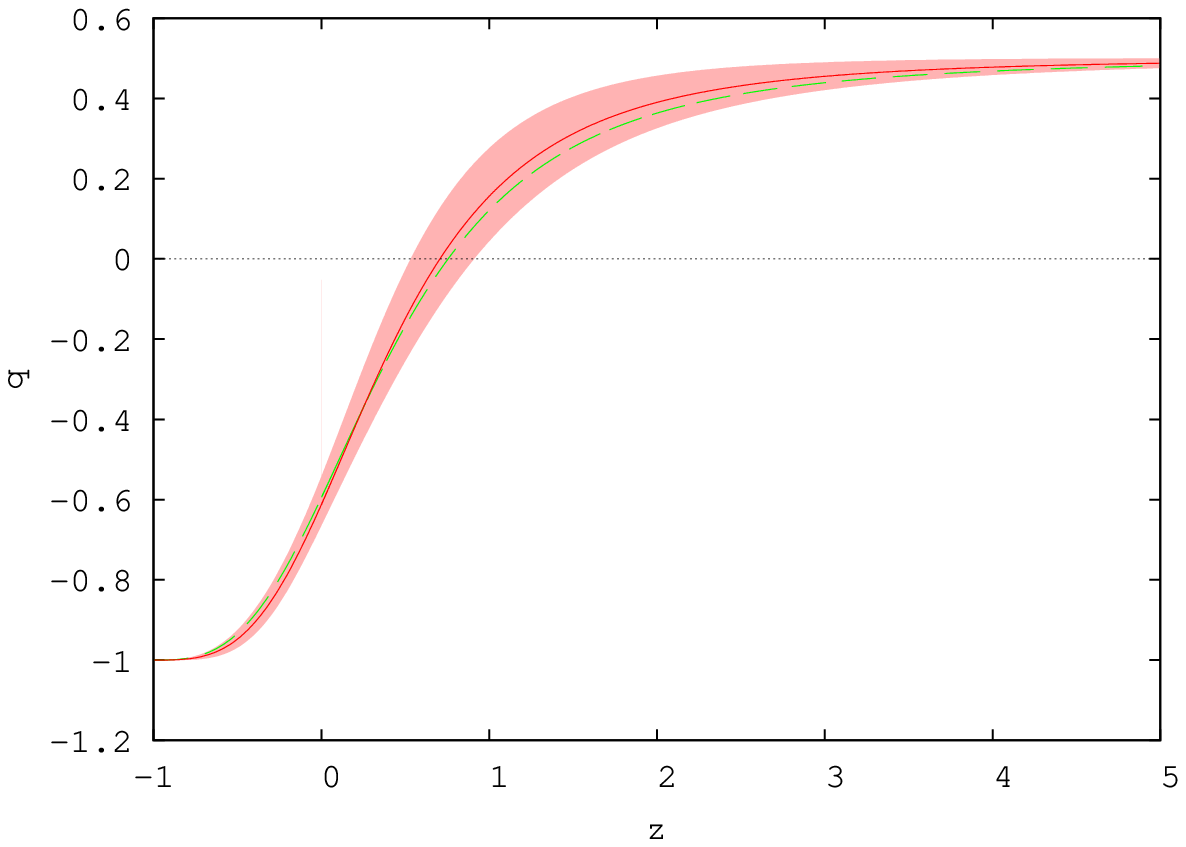}}&
      \resizebox{80mm}{!}{\includegraphics{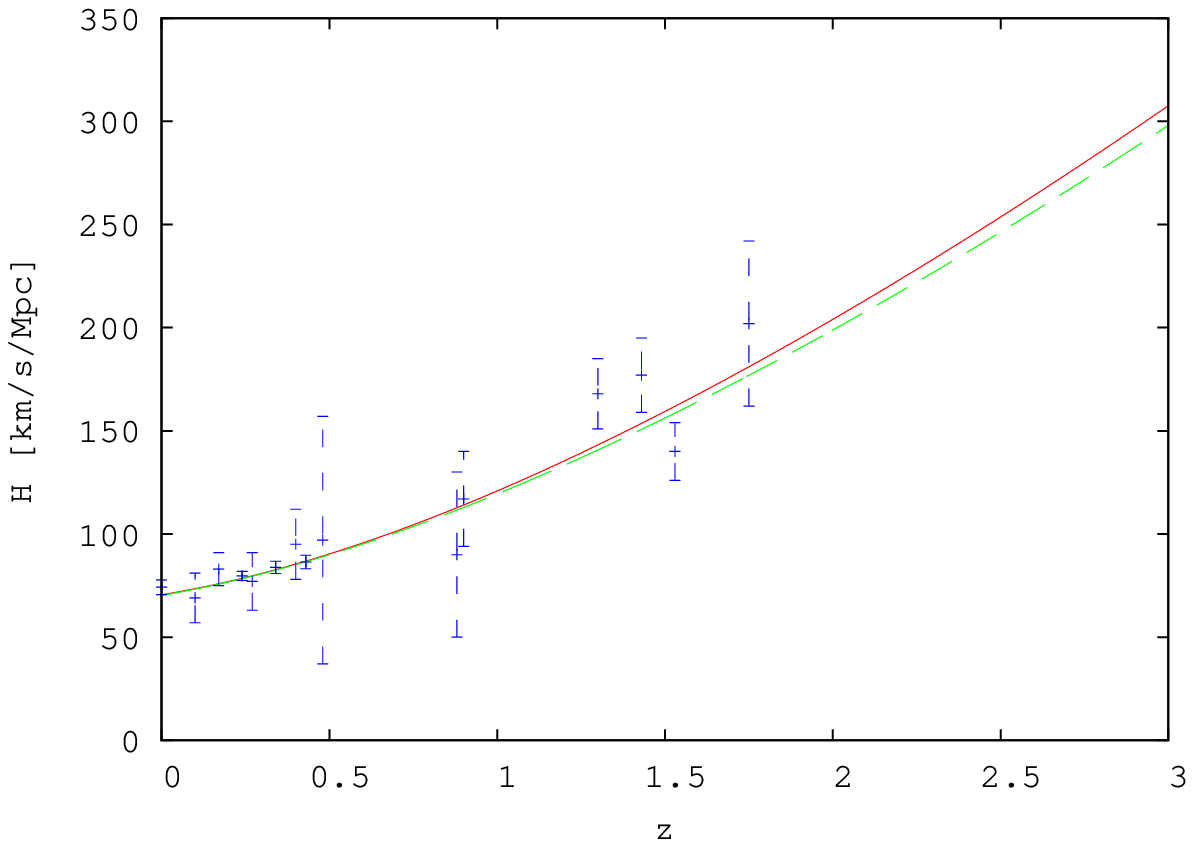}}\\
      \\
    \end{tabular}
    \caption{Left panel: deceleration parameter  vs. redshift. The
    shaded area shows the $1 \sigma$ confidence region.
    Right panel: Hubble function vs. redshift (the observational
    data are borrowed from Refs. \cite{Riess-2009,gazta,simon,stern}).
    In both panels the solid (red) and the dashed (green) lines are
    used for the best fit of parametrization I (Eq. (\ref{eq:q1})) and for the  $\Lambda$CDM model with
    $\Omega_{M0} = 0.27$ and $H_{0} = 72.1  \, {\rm km/s/Mpc}$ -see \cite{komatsu}-, respectively.
    The latter graph is shown for comparison.}
    \label{fig:h_q_1}
 \end{center}
\end{figure}

Figure \ref{fig:h_q_1} shows the evolution of the $q$ for the best
fit values of parametrization I (solid line), with its $1\sigma$
confidence region (shadowed area), and the spatially flat
$\Lambda$CDM model (dashed line) as determined by  the WMAP
7-years team \cite{komatsu} (the latter graph is included for the
sake of comparison), in the interval $-1 \leq z \leq 5$ (left
panel), and the evolution of the Hubble function in the interval
$0 \leq z \leq 3$ (right panel).

\begin{figure}[!htb]
  \begin{center}
    \begin{tabular}{cc}
      \resizebox{80mm}{!}{\includegraphics{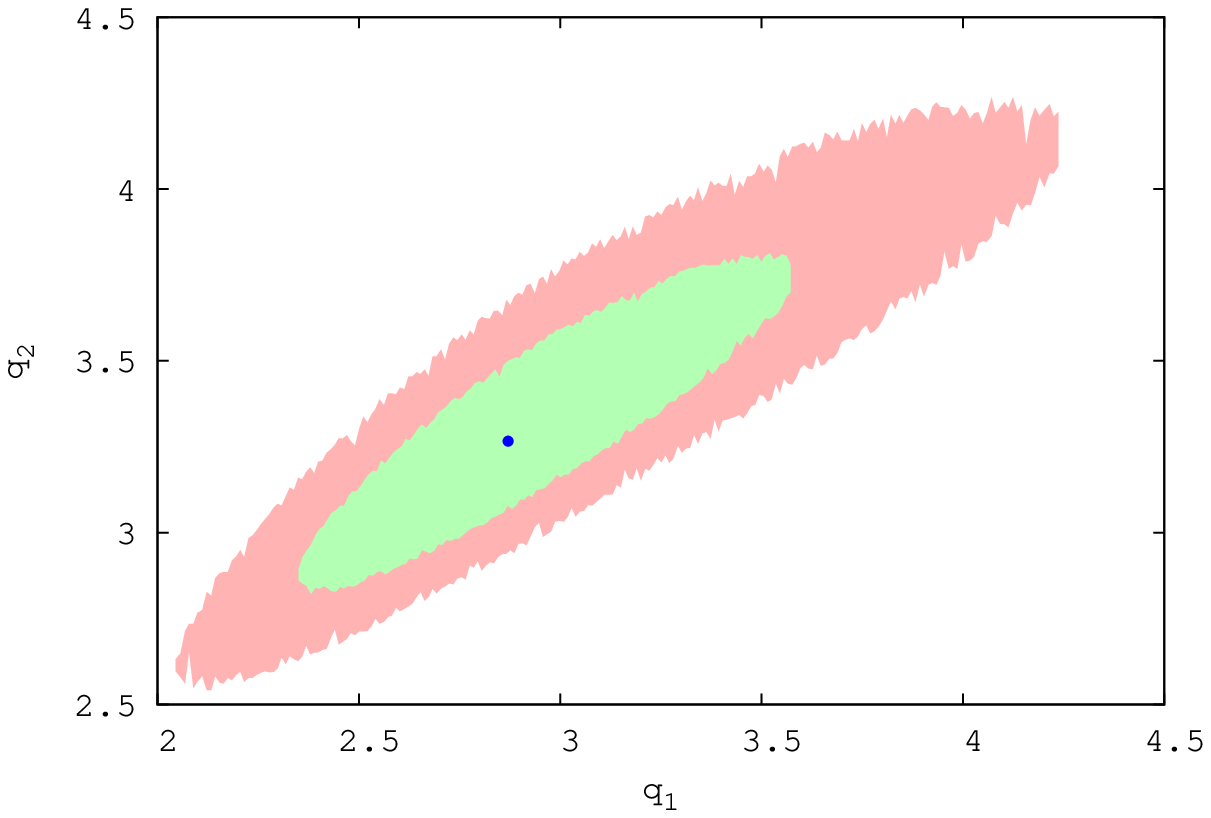}}&
      \resizebox{80mm}{!}{\includegraphics{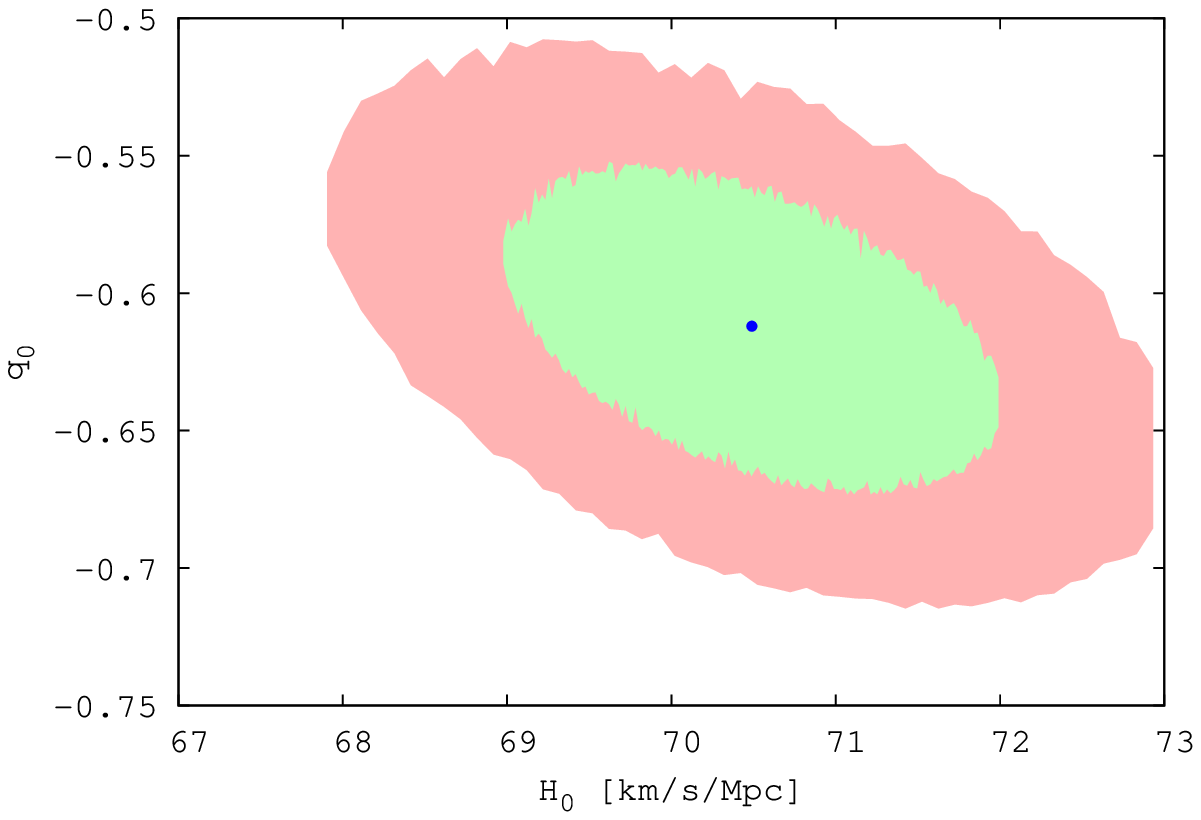}}\\
    \end{tabular}
    \caption{Left panel: 1$\sigma$ and 2$\sigma$ confidence regions of the pair of free parameters ($q_{1}, q_{2}$)
    of parametrization I, Eq. (\ref{eq:q1}). Right panel: 1$\sigma$ and 2$\sigma$ regions of
    the pair of free parameters ($H_{0}, q_{0}$). The dot signals the best fit values.}
    \label{fig:elipse_1}
 \end{center}
\end{figure}

Figure \ref{fig:elipse_1} depicts the $1 \sigma$ and $2 \sigma$
contour plots of the pairs ($q_{1}$, $q_{2}$) (left panel) and
($H_{0}$, $q_{0}$) (right panel).
Upon assuming that the expansion is dominated by pressureless
matter and some other (unspecified) component, non-interacting
between each other except gravitationally, the effective equation
of state (EoS) parameter is given by
\begin{equation}\label{eq:wdeq}
w(z)=\frac{2q(z)-1}{3(1-\Omega_{M}(z))}\, .
\end{equation}
Parametrizations of $q(z)$ and $w(z)$ are somewhat equivalent but
not quite because in the latter case some assumptions about the
energy budget of the Universe, as well on the existence or not of
possible interactions between the different components, have to be
made while in the former (as in our case) not necessarily.

Having said this, it is interesting to  confront (\ref{eq:wdeq})
with the widely used Chevallier-Polarsky-Linder (CPL)
parametrization \cite{chevallier,erik}
\begin{equation}\label{eq:CPL}
w=w_{0}+w_{1}\frac{z}{1+z} \, ,
\end{equation}
not far from $z=0$, in the redshift range $-0.3\leq z \leq 0.3$.
We restrict ourselves to comparative small redshift around $z=0$
because it diverges at $z\rightarrow -1$. After numerically
linearizing our expression for $w(z)$ we get $w_{0}=-0.92 \pm
0.10$ and $w_{1}=0.31^{+0.28}_{-0.25}$, values in very good
agreement with those reported in \cite{komatsu}, for the CPL
parameters, namely: $w_{0}=-0.93 \pm 0.12$ and
$w_{1}=-0.38^{+0.66}_{-0.65}$.

\begin{figure}[!htb]
  \begin{center}
    \begin{tabular}{c}
      \resizebox{120mm}{!}{\includegraphics{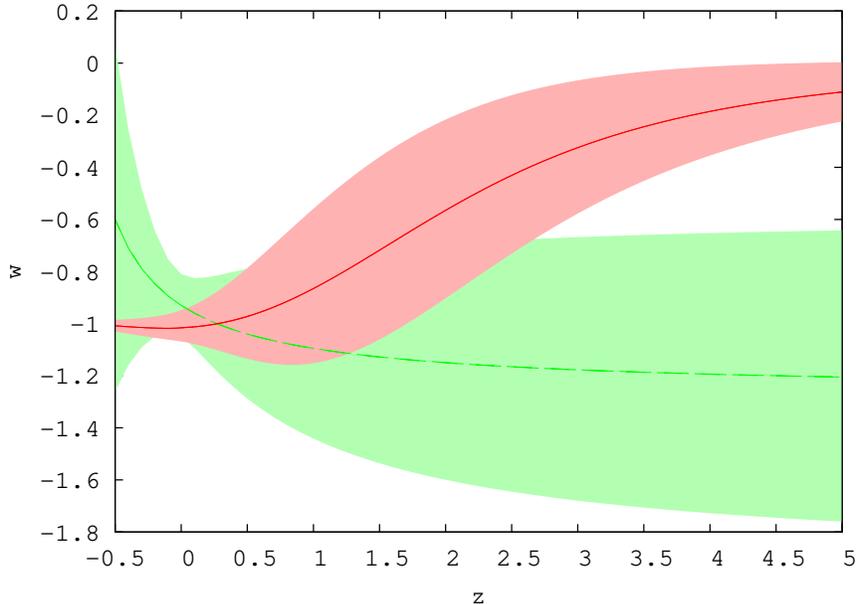}}\\
    \end{tabular}
    \caption{EoS parameters  vs. redshift. The
    shaded area shows the $1 \sigma$ confidence region.
    The solid (red) and the dashed (green) lines are
    used for the best fit of parametrization I (Eq. (\ref{eq:q1})) and for the
    CPL parameterization, Eq. (\ref{eq:CPL}), both with
    $\Omega_{M0} = 0.27 \pm 0.03$. For the CPL parameterization the values
    $w_{0}=-0.93 \pm 0.12$ and $w_{1}=-0.38^{+0.66}_{-0.65}$ obtained
    by Komatsu \textit{et al.} \cite{komatsu} were used.}
    \label{fig:w_1}
 \end{center}
\end{figure}

As Fig. \ref{fig:w_1} reveals, comparison in the extended interval
$-0.5 \leq z \leq 5$ shows that for $z \geq 2.5$ the evolution of
the effective of both EoS disagree in excess of $1 \sigma$. This
is consistent with claims that the CPL parametrization is not
appropriate to fit data simultaneously at low and high redshifts
\cite{li-wu-yu,cardenas}.

\subsection{Parametrization II}
As a second parametrization we propose,
\begin{equation}
 q(z) = -\frac{1}{4}\left(3q_{1}+1-3(q_{1}+1)\frac{q_{1}e^{q_{2}(1+z)}-e^{-q_{2}(1+z)}}
{q_{1}e^{q_{2}(1+z)}+e^{-q_{2}(1+z)}}\right) \, . \label{eq:q2}
\end{equation}

In this case  the Hubble function  must be obtained by numerically
integrating Eq. (\ref{eq:H(z)general}). Proceeding as before we
obtain, $q_{1}=0.078^{+0.086}_{-0.043} $,
$q_{2}=0.95^{+0.23}_{-0.20}$, and $H_{0}=70.4 \pm 1.6$
km/s/Mpc for the three parameters entering $H(z)$. The $\chi^{2}$
best fit values are shown in table \ref{tab:param2}.

\begin{figure}[!htb]
  \begin{center}
    \begin{tabular}{cc}
      \resizebox{80mm}{!}{\includegraphics{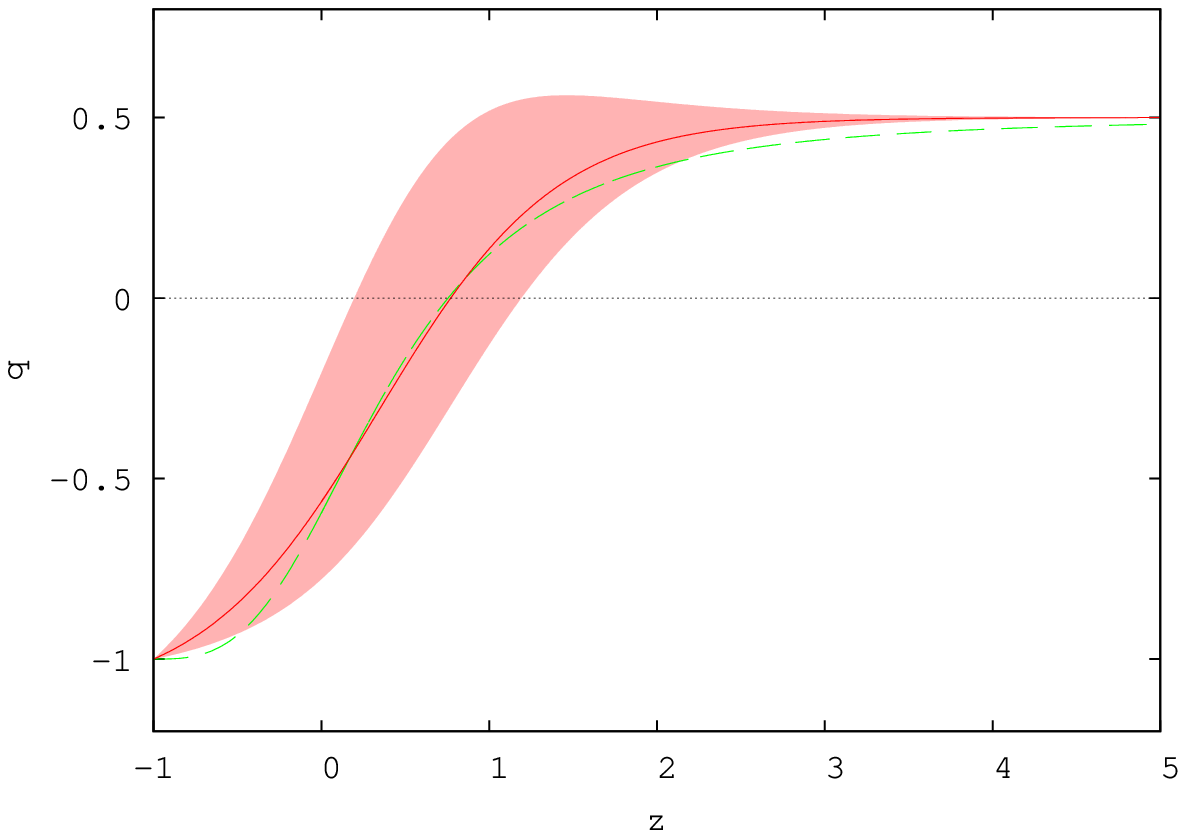}}&
      \resizebox{80mm}{!}{\includegraphics{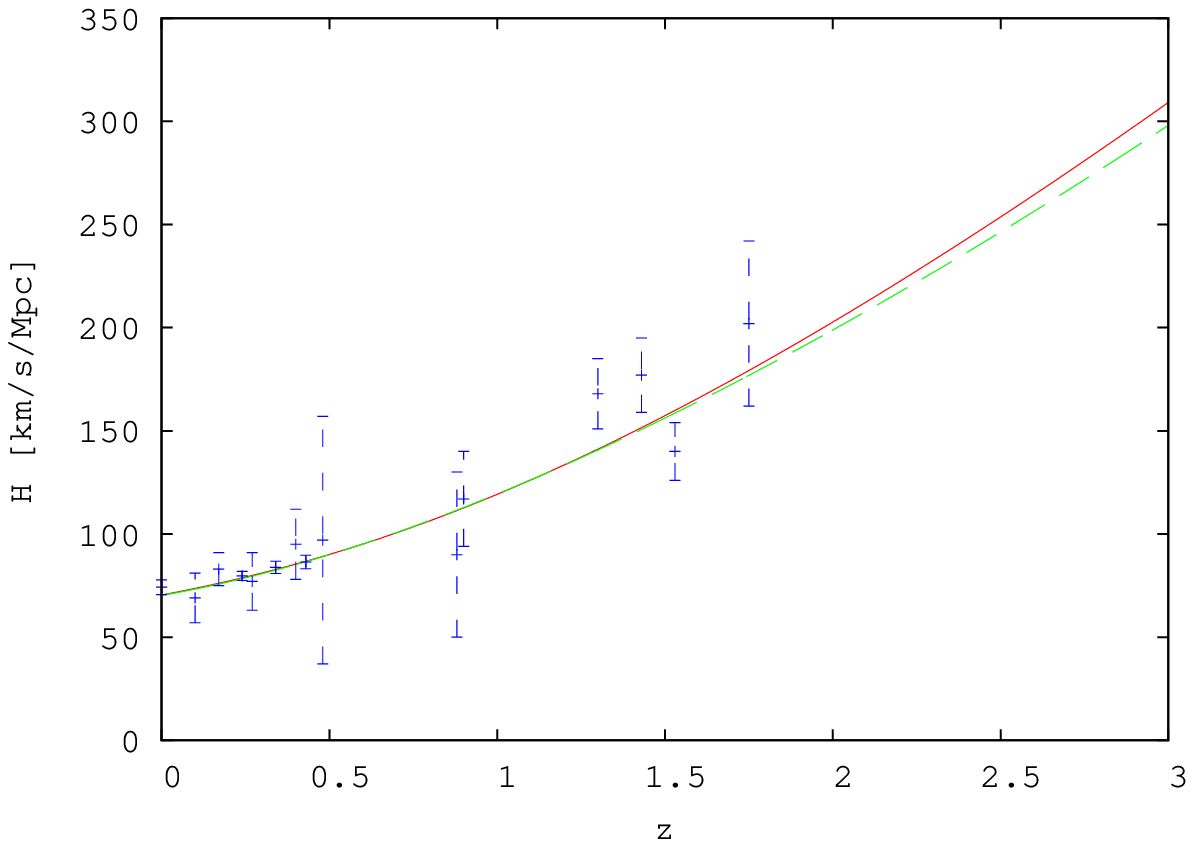}}\\
    \end{tabular}
    \caption{Same as Fig. \ref{fig:h_q_1} but for parametrization II, Eq. (\ref{eq:q2}).}
    \label{fig:h_q_2}
 \end{center}
\end{figure}

Left panel of Fig. \ref{fig:h_q_2}  shows the evolution of  $q$
for the best fit value of parametrization II (with its $1 \sigma$
confidence region) and the $\Lambda$CDM model obtained by Komatsu
{\it et al.} \cite{komatsu}. The right panel depicts the evolution
of the Hubble function versus redshift in the interval $0 \leq z
\leq 3$.
\begin{table}[!htb]
\begin{tabular}{||p{5 cm} || p{1.5 cm} p{1.8 cm} p{1.8 cm} p{1.8 cm} p{1.8 cm}||}
    \hline \hline
    Data sets &$\; \chi^{2}_{SN}$ &  $\chi^{2}_{BAO/CMB}$&  $\quad \; \chi^{2}_{H}$ & $\; \chi^{2}_{{\rm tot}}$ & $\chi^{2}_{{\rm tot}}/dof$ \\
    \hline \hline
    \footnotesize{Union2+CMB/BAO+Hubble} &$\; 542.7$  & $\quad \; \;  2.7$ & $\quad \; 17.7$ & $563.1$& $\; \; 0.96$ \\
    \hline \hline
    \end{tabular}
    \caption{Same as Table II but for  parametrization II, Eq. (\ref{eq:q2})}.
    \label{tab:param2}
\end{table}

\begin{figure}[!htb]
  \begin{center}
    \begin{tabular}{cc}
      \resizebox{80mm}{!}{\includegraphics{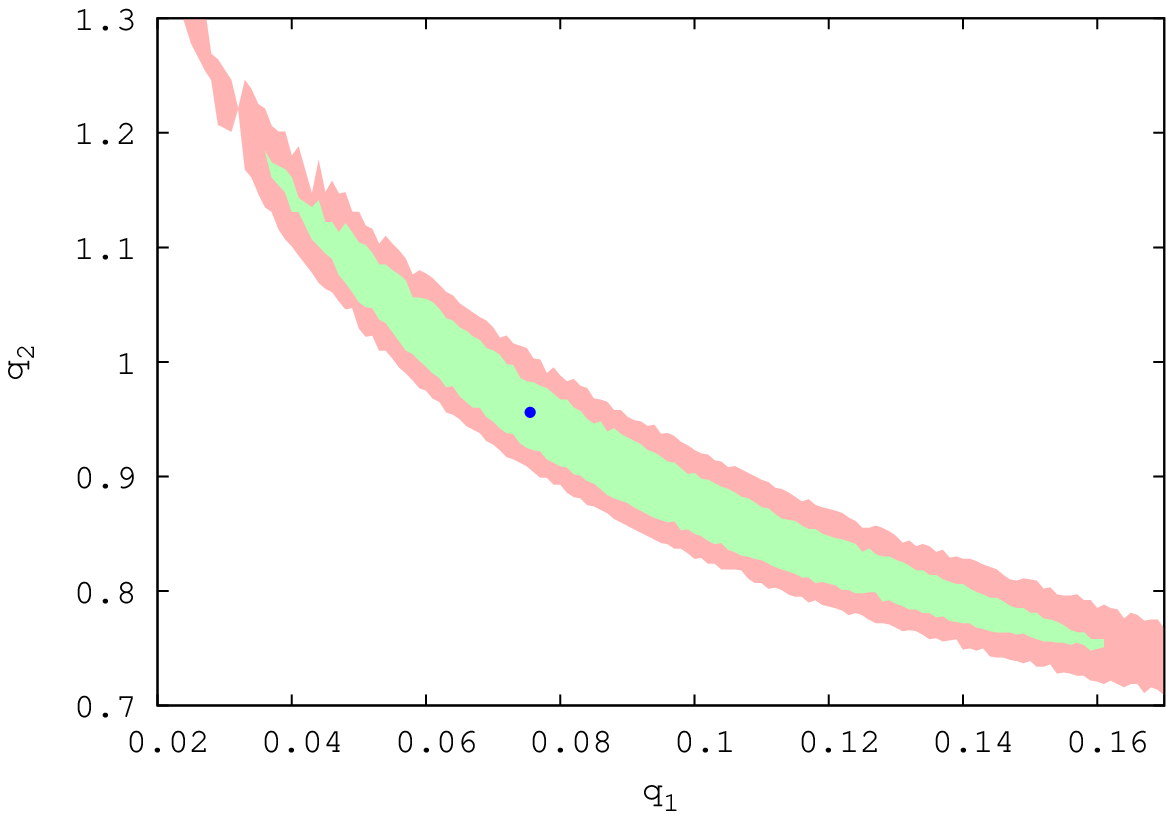}}&
      \resizebox{80mm}{!}{\includegraphics{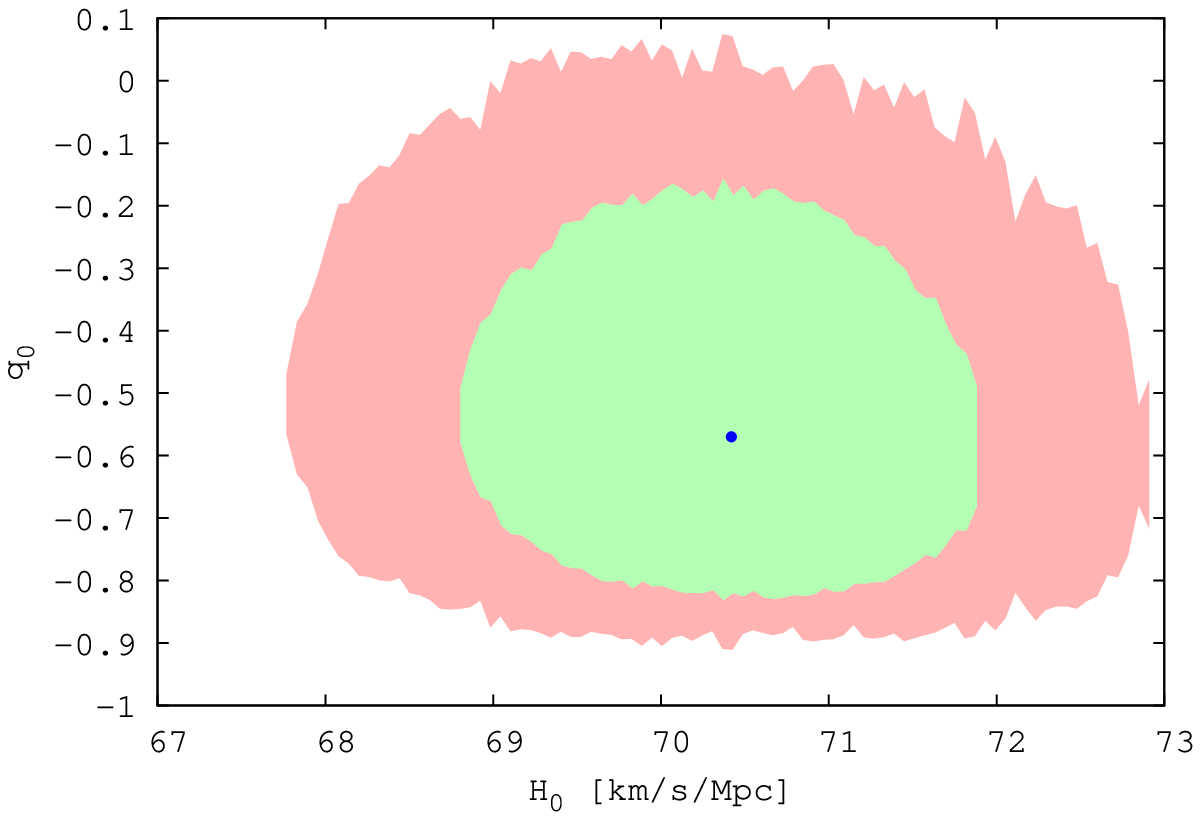}}\\
    \end{tabular}
    \caption{Same as Fig. \ref{fig:elipse_1} but for parametrization II, Eq. (\ref{eq:q2}).}
    \label{fig:elipse_2}
 \end{center}
\end{figure}

Figure \ref{fig:elipse_2} shows the $1 \sigma$ and $2 \sigma$
contour plots  of the pair ($q_{1}$, $q_{2}$) (left panel), and
($H_{0}$, $q_{0}$) (right panel). Note that $q_{0}$ results more
degenerate than in the previous parametrization (as well as in the
next one). This arises because -as direct inspection shows- in the
other two parametrizations  $q_{0}$ depends on just one free
parameter, $q_{1}$, while in this parametrization it depends on
both, $q_{1}$ and $q_{2}$.

\begin{figure}[!htb]
  \begin{center}
    \begin{tabular}{c}
      \resizebox{120mm}{!}{\includegraphics{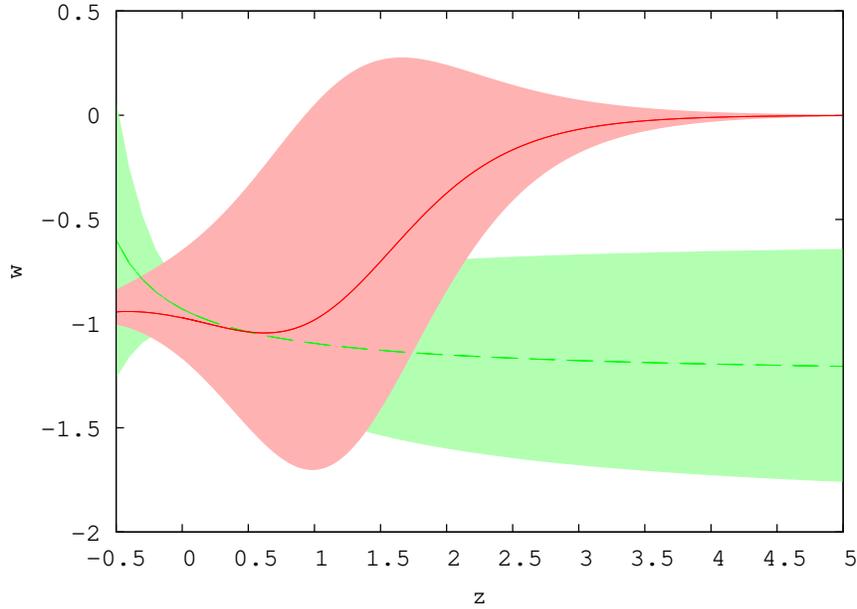}}\\
    \end{tabular}
    \caption{EoS parameters  vs. redshift. The
    shaded area shows the $1 \sigma$ confidence region.
    The solid (red) and the dashed (green) lines are
    used for the best fit of parametrization I (Eq. (\ref{eq:q1})) and  the
    CPL parametrization, Eq. (\ref{eq:CPL}), respectively, both
    assuming $\Omega_{M0} = 0.27 \pm 0.03$. For the CPL parametrization the values
    $w_{0}=-0.93 \pm 0.12$ and $w_{1}=-0.38^{+0.66}_{-0.65}$ obtained
    by Komatsu \textit{et al.} \cite{komatsu} were used.}
    \label{fig:w_2}
 \end{center}
\end{figure}

Considering the effective EoS parameter (\ref{eq:wdeq}), as in the
previous section, we obtain after linearization $w_{0} =
-0.97^{+0.33}_{-0.21}$ and $w_{1} = -0.15^{+0.70}_{-0.47} $. As
Fig. \ref{fig:w_2} shows, the evolution of the said effective EoS
and the CPL in an extended redshift interval is similar to the
previous one but with the $1\sigma$ uncertainty interval
significantly wider.

\subsection{Parametrization III}
The previous parametrization presents the inconvenience of a
significant uncertainty in $q_{0}$ since it depends on the two
free parameters. The following parametrization
\begin{equation}
q(z) = -\frac{1}{4} \, + \, \frac{3}{4} \; \frac{q_{1}
e^{q_{2}\frac{z}{\sqrt{1+z}}}-e^{-q_{2}
\frac{z}{\sqrt{1+z}}}}{q_{1}
e^{q_{2}\frac{z}{\sqrt{1+z}}}+e^{-q_{2}\frac{z}{\sqrt{1+z}}}} \, .
\label{eq:q3}
\end{equation}
avoids this as $q_{0}$ depends on $q_{1}$ only.

\begin{figure}[!htb]
  \begin{center}
    \begin{tabular}{cc}
      \resizebox{80mm}{!}{\includegraphics{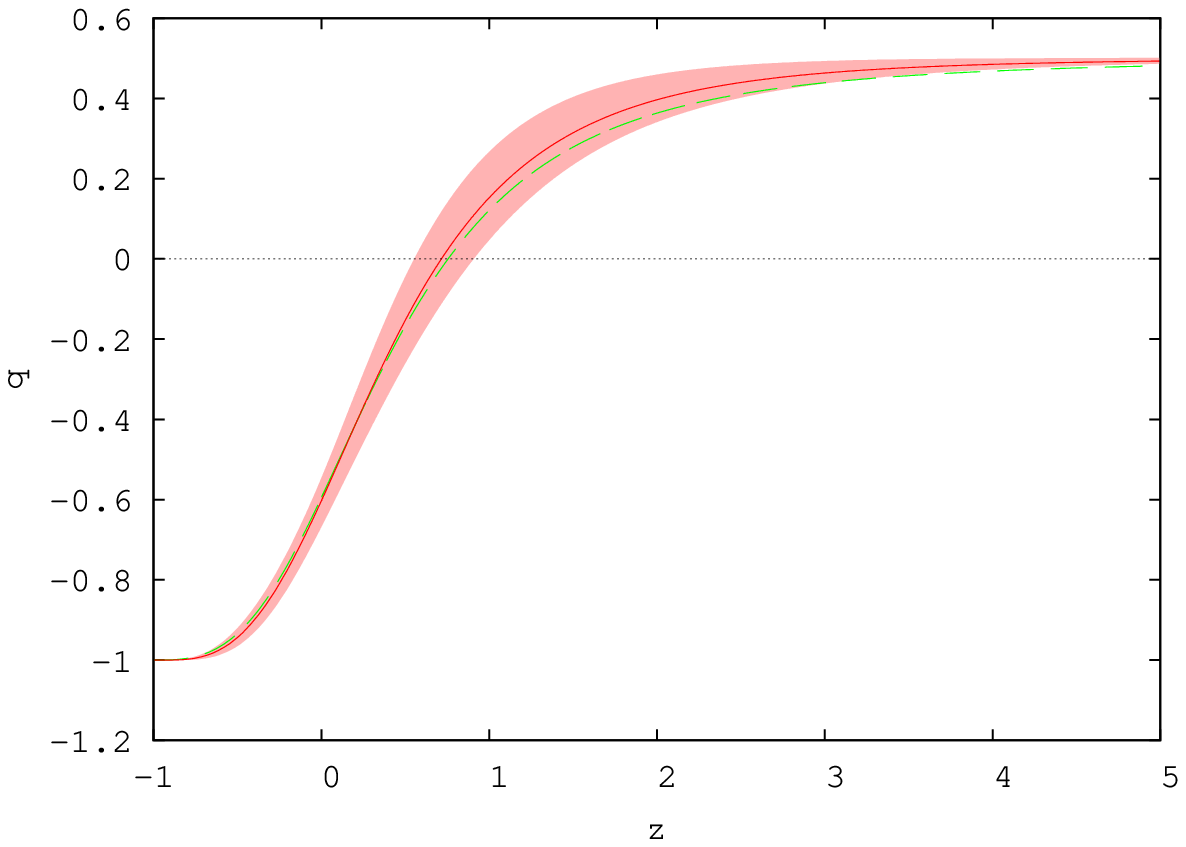}}&
      \resizebox{80mm}{!}{\includegraphics{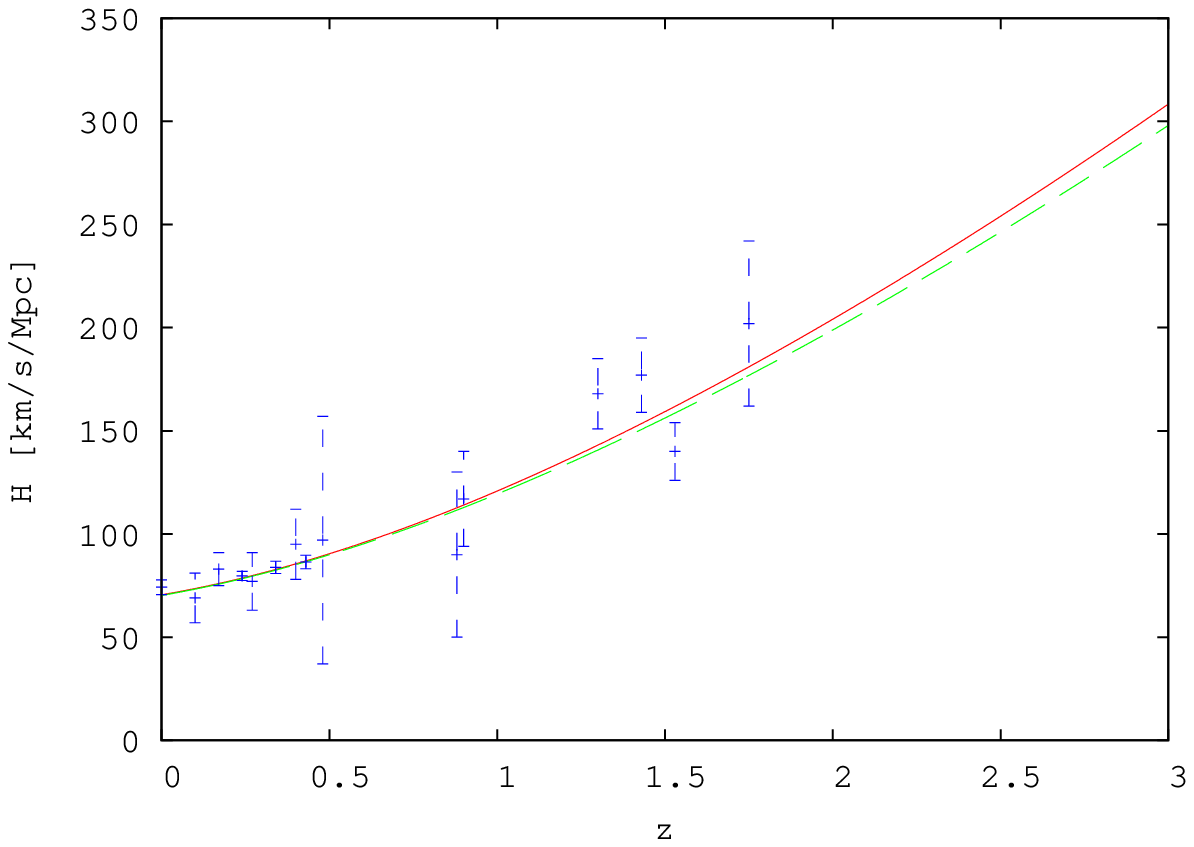}}\\
      \\
    \end{tabular}
    \caption{Same as Fig. \ref{fig:h_q_1}  but for parametrization III, Eq. (\ref{eq:q3}).}
    \label{fig:h_q_3}
 \end{center}
\end{figure}

Again, $H(z)$ must be obtained numerically. Then, proceeding as in
the two previous instances, we obtain $q_{1}=0.36^{+0.07}_{-0.08}
$, $q_{2}=1.57^{+0.27}_{-0.33}$, and $H_{0}=70.5^{+1.4}_{-1.6}$
km/s/Mpc. The $\chi^{2}$ values of the best the fit are indicated
in table \ref{tab:param3}.
\begin{table}[!htb]
    \begin{tabular}{||p{5 cm} || p{1.5 cm} p{1.8 cm} p{1.8 cm} p{1.8 cm} p{1.8 cm}||}
    \hline \hline
    Data sets &$\; \chi^{2}_{SN}$ &  $\chi^{2}_{BAO/CMB}$&  $\quad \; \chi^{2}_{H}$ & $\; \chi^{2}_{{\rm tot}}$ & $\chi^{2}_{{\rm tot}}/dof$ \\
    \hline \hline
    \footnotesize{Union2+CMB/BAO+Hubble} &$\; 542.6$  & $\quad \; \;  1.7$ & $\quad \; 17.9$ & $563.2$& $\; \; 0.96$ \\
    \hline \hline
    \end{tabular}
    \caption{Same as Table II but for parametrization III, Eq. (\ref{eq:q3}).}

    \label{tab:param3}
\end{table}
Figure \ref{fig:h_q_3} shows the evolution of the deceleration
parameter  for the best fit parametrization (with its $1\sigma$
confidence region) and the $\Lambda$CDM model as determined the
WMAP team \cite{komatsu} (left panel), and the evolution of the
Hubble function (right panel).

\begin{figure}[!htb]
  \begin{center}
    \begin{tabular}{cc}
      \resizebox{80mm}{!}{\includegraphics{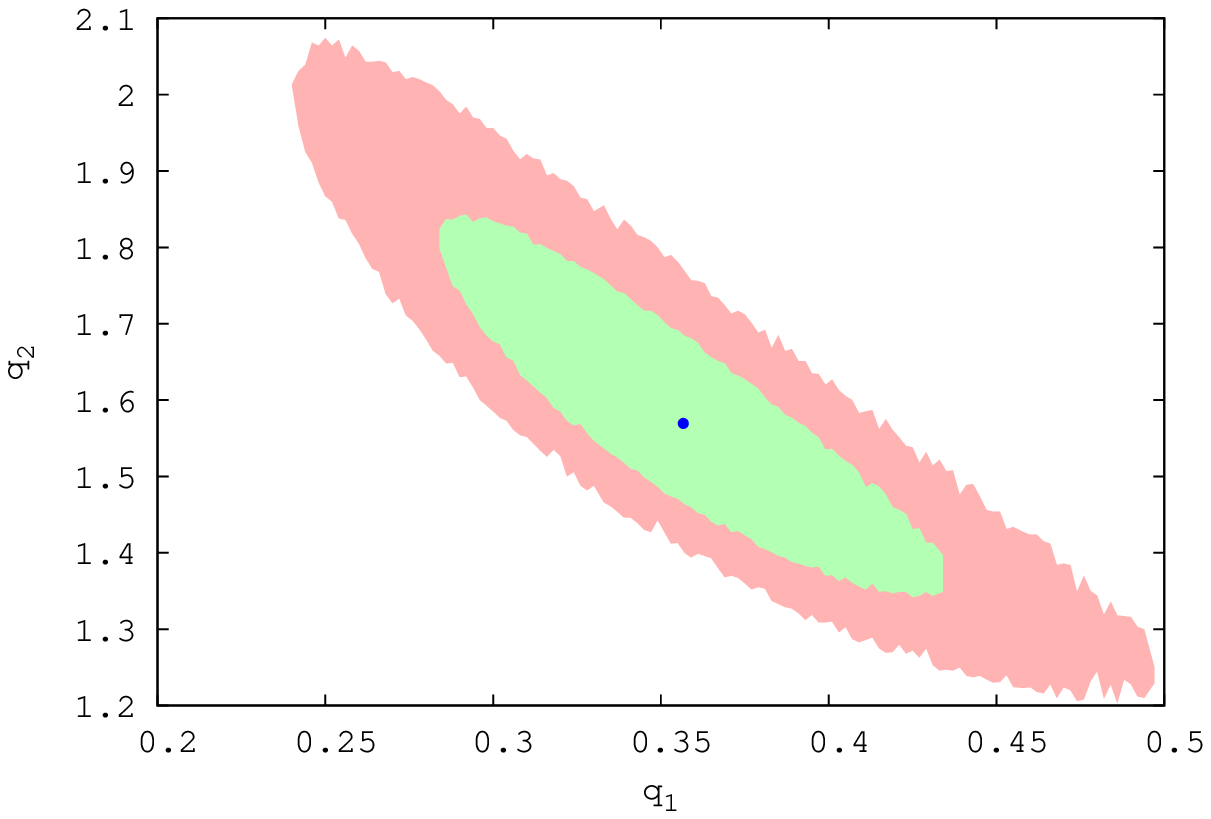}}&
      \resizebox{80mm}{!}{\includegraphics{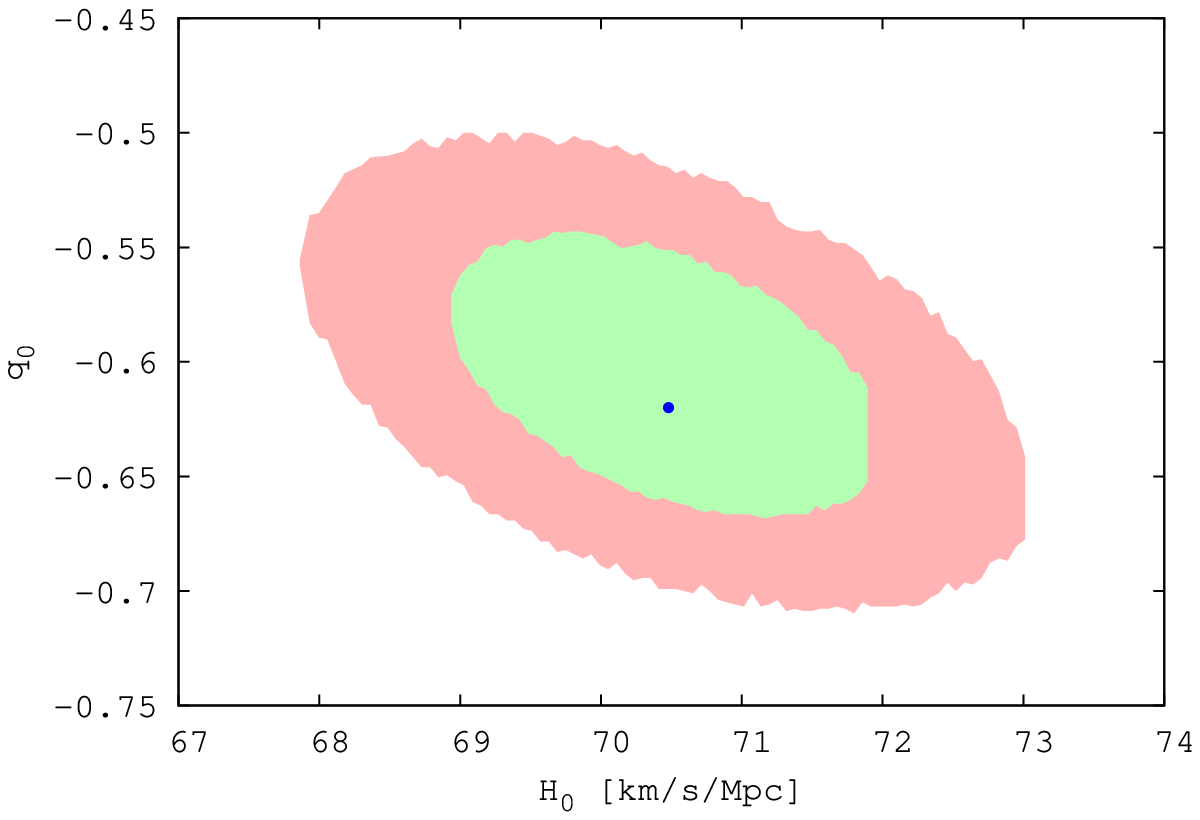}}\\
    \end{tabular}
    \caption{ Same as Fig. \ref{fig:elipse_1}  but for parametrization III, Eq. (\ref{eq:q3}).}
    \label{fig:elipse_3}
 \end{center}
\end{figure}

Fig. \ref{fig:elipse_3} shows the $1 \sigma$ and $2 \sigma$
contour plots of the pairs ($q_{1}$, $q_{2}$) (left panel) and
($H_{0}$, $q_{0}$) (right panel).

As for the effective EoS parameter (\ref{eq:wdeq}), proceeding as
in the previous subsections, we obtain $w_{0}=-1.01 \pm 0.06$ and
$w_{1}=0.03\pm 0.16$. Again, the evolution of the said effective
EoS and the CPL in an extended redshift interval is rather similar
to the one in Fig. \ref{fig:w_1}; thereby, we do not feel it
necessary to show it here.
\subsection{Discussion}
\begin{figure}[!htb]
  \begin{center}
    \begin{tabular}{c}
    \resizebox{120mm}{!}{\includegraphics[width=12cm]{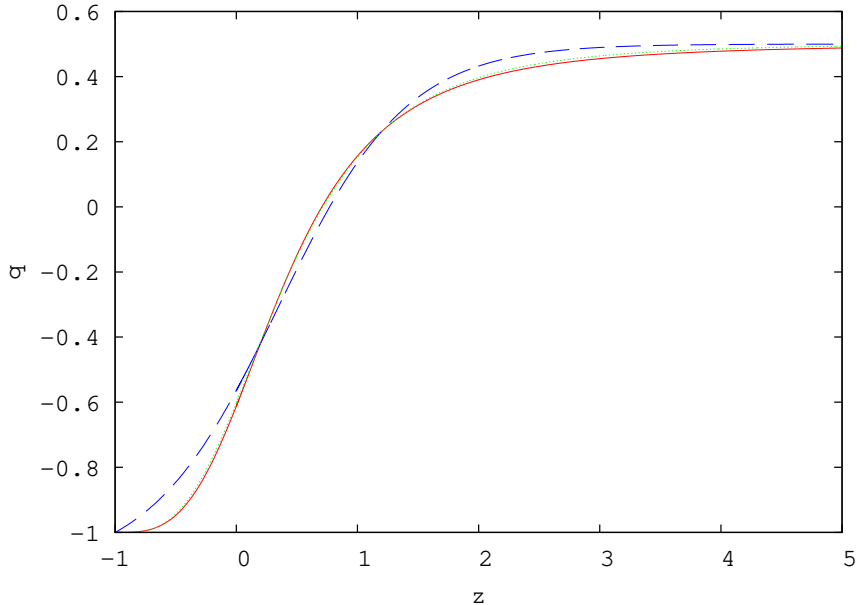}}
    \end{tabular}
\caption{Deceleration parameters vs. redshift. Solid (red), long
dashed (blue) and short dashed (green) lines are for
parametrizations I, II and III respectively. The graphs of
parameterizations I and III practically overlap each other.}
\label{fig:qs_z}
 \end{center}
\end{figure}

Figure \ref{fig:qs_z} compares the parametrizations. All three
yield rather similar results being really close between one
another from the statistical standpoint ($\chi^{2}/dof =0.96$ for
all of them). However, parametrization II looks somewhat less
favored than the other two because of the noticeably wider
$1\sigma$ region of $q$ vs. $z$, as seen in the left panel of Fig.
\ref{fig:h_q_2}.

All of them are consistent with a present stage of accelerated
expansion, never to end or  slow down. Further, the best fit plots
of $H(z)$ and $q(z)$ are alike the corresponding plots of the
$\Lambda$CDM model as determined by Komatsu {\it et al.}
\cite{komatsu}. From Table \ref{tab:Hoqoz} we learn that all
$H_{0}$ best fit values are within $1 \sigma$ of each other and
consistent with the $H_{0}$ value reported in \cite{komatsu}. The
same holds true for the best fits of the age of the Universe
(given by $t_{0} = \int_{0}^{\infty}{dz/[(1+z)\, H(z)]}$), $q_{0}$
and the redshift, $z_{t}$, at which the transition
deceleration-acceleration occurred (i.e., $q(z_{t}) = 0$), though
the central values of the latter are not so close between each
other as the corresponding values of the other two parameters. At
any rate, the three of them are consistent with the  $z_{t}
\approx 0.5$ value obtained by Wu {\it et al.} using the history
of the strong energy condition \cite{wumazhang}, as well as with
the findings of Riess {\it et al.} \cite{adam2007}, Cunha and Lima
\cite{ademir}, and Lu {\it et al.} \cite{lu-xu}.

As table \ref{tab:Hoqoz} shows, the values predicted for Hubble's
constant, $H_{0}$, by the three parametrizations are within
$1\sigma$ between one another and with the value predicted by the
$\Lambda$CDM model that best fit identical sets of observational
data. This is also true for $t_{0}$, $q_{0}$ and $z_{t}$.

\begin{table}[!htb]
     \begin{tabular}{||p{1.2 cm}|| p{2.6 cm} p{2.6 cm} p{2.4 cm} p{2.7 cm}||}
\hline \hline
 & Param. I & Param. II & Param. III &$\; \;\;\;\;  \Lambda$CDM \\
\hline \hline $\; \; H_{0} $ &$\; \, 70.5^{+1.5}_{-1.6} $ & $\; \,
70.4 \pm 1.6$ & $\; \, 70.5^{+1.4}_{-1.6}$&
$\; \; \; \; 70.2 \pm 1.4 $ \\
\hline $\; \; t_{0} $ &$\; \, 13.6 \pm 0.5$ & $\; \, 13.7\pm 0.4$
& $\; \, \, 13.6\pm 0.2$& $\; \; \; \; 13.4\pm 0.1$ \\
\hline
 $\; \; q_{0} $ &$\; -0.61^{+0.06}_{-0.07}$ & $\; -0.56^{+0.35}_{-0.22}$& $\;-0.60\pm 0.06$ &
 $\; \, -0.60 \pm0.03$ \\
\hline $\; \; z_{t}$ & $\; \; \; 0.71^{+0.14}_{-0.17}$ &$\;  \;
\;\, 0.77^{+0.52}_{-0.57} $ & $\;  \; \;\, 0.72^{+0.27}_{-0.21}$ &
$\; \; \; \; \; 0.76 \pm 0.05$ \\ \hline \hline
\end{tabular}
\caption{Hubble's constant, $H_{0}$ (in km/s/Mpc), the age of the
Universe, $t_{0}$  (in Gyr), the deceleration parameter, $q_{0}$,
and the redshift, $z_{t}$, of the transition
deceleration-acceleration for the three parametrizations, and the
flat $\Lambda$CDM model fitted to the same data sets.}
\label{tab:Hoqoz}
\end{table}

A direct and model-independent determination  of $q(z)$ in the
redshift interval $0 \leq z \leq 1$ was carried out by Daly  and
coworkers \cite{daly1} who applied the expression \cite{daly_2005}
\begin{equation}
q(z) = - 1 \, - \,(1+z)\, \left[\frac{d^{2}y/dz^{2}}{dy/dz} \, +
\, \frac{\Omega_{k0} \, y \, dy/dz}{1\, +\, \Omega_{k0}
y^{2}}\right] \label{eq:q(z)daly}
\end{equation}
to the 192 SN I data points of Davis {\it et al.} \cite{davis} and
30 radiogalaxy data points of Daly  {\it et al.} \cite{daly2} -see
Fig. 10 in \cite{daly1}. In (\ref{eq:q(z)daly}) $ y(z) = H_{0}
(a_{0} \, r)$ is the dimensionless coordinate distance, $r$  the
radial coordinate of the FLRW metric, and $ \Omega_{k0} = -
k/(H_{0}\, a_{0})^{2}$. Notice that it just assumes the FLRW
metric; i.e., it holds regardless the energy components of the
Universe or the specific theory of gravity adopted.

\begin{figure}[!htb]
  \begin{center}
    \begin{tabular}{c}
    \resizebox{180mm}{!}{\includegraphics[width=16cm]{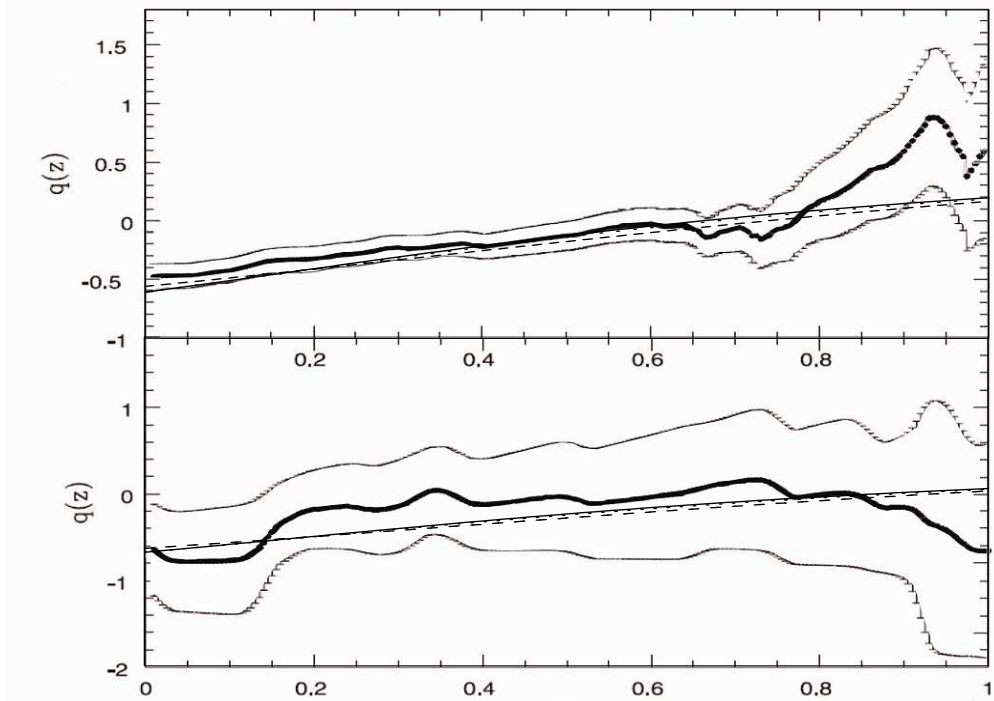}}
    \end{tabular}
\caption{Top panel: observational constraints, at $1\sigma$
confidence level, on the deceleration parameter versus redshift,
obtained by Daly {\it et al.} using the combined sample $192$
supernovae and $30$ radio galaxies (top panel of Fig. 10 in
\cite{daly1}); the thick solid line corresponds to the central
experimental value of $q(z)$. Solid, dashed, and dot-dashed lines
correspond to parametrizations I (Eq. (\ref{eq:q1})), II (Eq.
(\ref{eq:q2})) and III (Eq. (\ref{eq:q3})), respectively. Bottom
panel: The same as the top panel except that the observational
constraints on $q(z)$ were obtained solely from the sample of $30$
radio galaxies (bottom panel of Fig. 10 in \cite{daly1}). In both
panels  parametrizations I and III practically overlap each
other.} \label{fig:daly}
 \end{center}
\end{figure}

Figure \ref{fig:daly} shows the best fit graphs of $q(z)$ of each
parametrization superimposed to experimental results of Daly {\it
et al.} \cite{daly1}. As it is seen, these lines fall within the
$1\sigma$ region of $q(z)$ as determined in Ref. \cite{daly1}.

One may ask as to what extent the imposing of the
thermodynamic-based, far future, constraint $q(z =-1) = -1$ biases
the parametrizations  toward the $\Lambda$CDM model. We have
studied this by letting the value of $q(z = -1)$ as an additional
free-parameter  and fitting it using solely the observational
data. The results are: $q (z=-1)=-0.65^{+0.21}_{-0.50}$, $q (z =
-1) =-1.1^{+0.2}_{-1.7}$, and $ q (z = -1)=-0.82^{+0.07}_{-0.4}$
for parametrizations I, II and III, respectively. Except for the
second one, the quintessence cold dark matter (CDM) model is
somewhat preferred but, in all the cases, the physically motivated
choice $q(z = -1) = -1$ results compatible within $1 \sigma$. The
drawback of letting $q(z = -1)$ free, aside from violating
thermodynamics, is that the other two free-parameters present a
wide degeneracy.

Likewise, the derived values for the Hubble constant (first row in
Table \ref{tab:Hoqoz}) differ from the recently obtained  by Riess
et al., $H_{0} = 74.2 \pm 3.6$ km/s/Mpc, \cite{Riess-2009}, who
used $240$ Cepheids variables at $z < 0.1$, by about $6\%$ (but
they all agree with the latter at $1 \sigma$). We have considered
this by repeating the analysis of subsection IIIC but this time
leaving aside the mentioned value of Riess et al. The results now
are: $H_0=70.0^{+1.7}_{-1.5}$, $H_0=70.0^{+1.5}_{-1.6}$, and
$H_0=70.1^{+1.6}_{-1.6}$ Km/s/Mpc for parametrizations I, II and
III, respectively. Thus, both sets of results are essentially
coincident (they differ by less than $1\%$). They also agree very
well with the Hubble constant value observationally derived by
Komatsu et al., $70.4 \pm 2.5$ km/s/Mpc \cite{komatsu}, using WMAP
7-year data. So, while there is a significant difference between
the Hubble constant value of Riess et al. and ours, it is not a
substantial one; after all, they agree at $1 \sigma$ confidence
level. At any rate, the root of the discrepancy may be rightly
traced at the difference in methods employed. While Riess et al.
essentially used astrophysical data, we (as well as Komatsu et
al.) resorted to cosmological data instead.

\section{Concluding remarks}\label{sec:remarks}
In this paper we proposed three different two-parameter
parametrizations of $q(z)$ valid from the matter era $(z \gg 1)$
up to the infinite future ($ z= -1$), modulo $H (z)> 0$. These
rest in the following hypotheses: $(i)$ at cosmological scales the
Universe is homogeneous and isotropic, thereby well described by
the FLRW metric; $(ii)$ in the matter dominated era $q = 0.5\, $;
$(iii)$ at least at late times the entropy of the Universe is
dominated by the entropy of the apparent horizon. The second an
third hypotheses furnish two fixed points (at $z \gg 1$ and $z =
-1$, respectively), thereby drastically reducing the ample
latitude one faces in parameterizing $q(z)$. By smoothly
interpolating between these two points one can obtain useful
parametrizations, but with shrunk arbitrariness.

Except for the existence of a matter dominated era at early times,
the parametrizations  are independent on any specific cosmological
model; and, on the other hand, they are flexible enough to
accommodate many homogeneous and isotropic models. We constrained
the free parameters with the latest observational data (SN Ia,
BAO, CMB, and $H(z)$). Accordingly if to accommodate a given
cosmological model the free parameters, $q_{1}$ and $q_{2}$, in
the three parametrizations should take values widely apart from
their respective best fits (which are consistent within $1\sigma$
with the flat $\Lambda$CDM model), we may confidently discard the
said model.

Thermodynamics in spatially  flat ($k = 0$) FLRW  universes
demands that $q(z = -1) = -1$. This provides us with an additional
and very useful fixed point to parametrize $q(z)$ in a model
independent manner. Note that in the absence of a physically-based
guidance to figure out the value of the deceleration parameter at
$z = -1$ one is led to choose some or other random value. By
contrast, in our case we have taken $q(z= -1) = -1$ on solid
thermodynamic grounds.

Albeit we have considered just the particular set of spatially
flat FLRW universes, it is not a big restriction at all. Indeed,
recalling that in the case of non-flat metrics  the area of the
apparent horizon is given by Eq. (\ref{SA}) it follows that ${\cal
A}' = {\cal A}^{2}/(2 \pi a)\, [H^{2} (1+q) \, + \, k \, a^{-2}]$.
When $a \rightarrow \infty$ and $k = -1$, the last term on the
right hand side is necessarily subdominant otherwise one would
have ${\cal A}' < 0$, contrary to the second law. Hence the
condition ${\cal A}' \geq 0$ in that limit reduces to the one in
flat space, namely, $q \geq -1$. In the positively curved case,
and again in the same limit, one can assume that $H \propto
a^{n}$, $n$ being some real number. As it can be straightforwardly
checked, the aforementioned last term results, once more,
subdominant provided that $n > -1$ which is the case of most
realistic cosmologies. Thus, $(z= -1, \, q =-1)$ is an asymptotic
fixed point not only for spatially flat universes but also for
open universes and for a rather ample set of closed universes.

Our results suggest that from the era of matter domination  onward
$q$ decreases monotonously with expansion (i.e., $dq/dz > 0$), the
transition deceleration acceleration occurred  at a redshift of
about $0.7$, and that $q_{0} \simeq -0.6$. They do not support
recent claims that the cosmic expansion is today reverting to a
decelerated phase (i.e., that $dq/dz|_{0} < 0$) \cite{cardenas},
\cite{shafieloo}-\cite{zuniga}. On the contrary, they show overall
consistency with the findings of \cite{elgaroy}-\cite{giostri},
\cite{daly1}, \cite{guimaraes-2009}, as well as with those of
Serra {\it et al.,} \cite{serra}. The latter authors showed that
the equation of state of dark energy has not varied noticeably in
the redshift interval $0 \leq z \leq 1$.

\newpage
\acknowledgments{We are indebted to Bin Wang for useful advice and
the anonymous referee for helpful comments. This work was
partially supported from ``Comisi\'{o}n Nacional de Ciencias y
Tecnolog\'{\i}a" (Chile) through the FONDECYT Grant No. 1110230
(SdC and RH) and No. 1090613 (RH and SdC). D.P. acknowledges
``FONDECYT-Concurso incentivo a la cooperi\'{o}n internacional"
No. 1110230 and is grateful to the ``Instituto de F\'{\i}sica",
where part of this work was done, for warm hospitality. Also I.D.
and D.P. research was partially supported by the ``Ministerio
Espa\~{n}ol de Educaci\'{o}n y Ciencia" under Grant No.
FIS2009-13370-C02-01, and by the ``Direcci\'{o} de Recerca de la
Generalitat" under Grant No. 2009SGR-00164.}

\end{document}